\documentclass[12pt]{article}

\usepackage{latexsym,amsfonts,amsmath,amsthm,amssymb,amsbsy,multirow ,textcomp,hyperref,wrapfig,datetime,verbatim,cancel,subfigure,cite}
\usepackage[dvipdfmx]{graphicx}
\usepackage[font={footnotesize,sl},bf]{caption}

\def\be{\begin{equation}}
\def\ee{\end{equation}}
\def\bea{\begin{eqnarray}}
\def\eea{\end{eqnarray}}
\newcommand{\beq}{\begin{eqnarray}}
\newcommand{\eeq}{\end{eqnarray}}

\newcommand{\ba}{\begin{align}}
\newcommand{\ea}{\end{align}}

\numberwithin{equation}{section}

\long\def\new#1\endnew{{\bf #1}}		
\long\def\del#1\enddel{}

\def\del{\partial}
\def\nn{\nonumber}
\usepackage{color}
\usepackage[dvipsnames]{xcolor}

\newcommand{\pink}[1]{\textcolor{\pink}{#1}}

\def\O{\mathcal{O}}

\def\tJ{\widetilde J}
\def\tr{\text{tr}}
\def\g{\mathfrak{g}}

\def\nn{\nonumber}

\def\O{\mathcal{O}}
\def\T{\mathcal{T}}
\def\R{\mathbb{R}}

\newcommand{\ket}[1]{\left| #1\right>}
\newcommand{\bra}[1]{\left< #1\right|}

\newcommand*\caliw{\includegraphics[scale=0.0139]{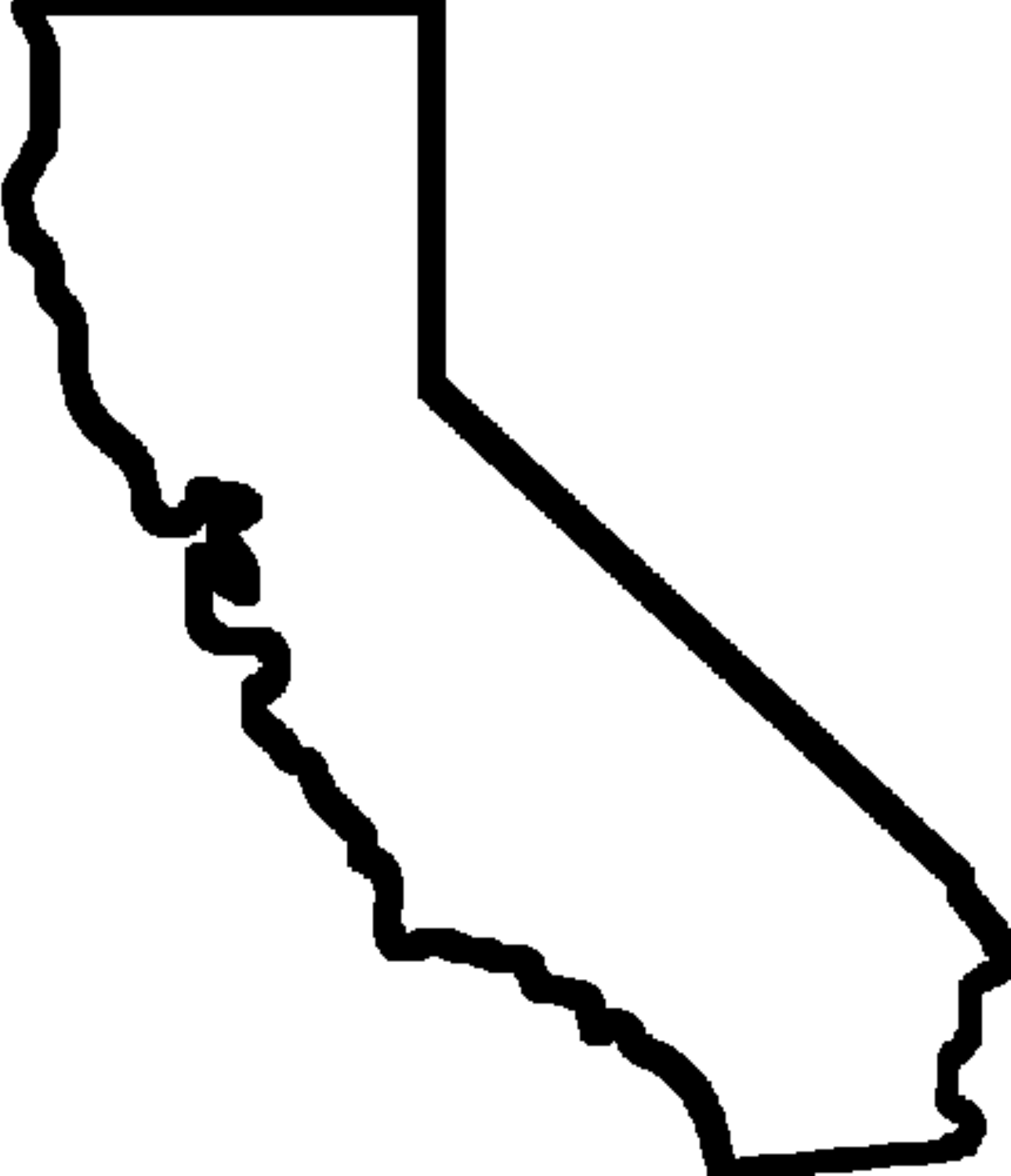}}
\newcommand*\calib{\includegraphics[scale=0.07]{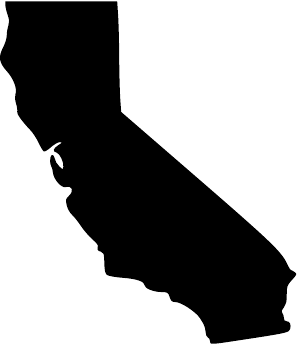}}

\begin{document}

 \begin{titlepage}
  \thispagestyle{empty}
  \begin{flushright}
  CPHT-RR102.122017
  \end{flushright}
  \bigskip
  \begin{center}
  \baselineskip=13pt {\LARGE Corrections in the relative entropy of black hole microstates}
   \vskip1.5cm
   \centerline{
   {\bf Ben Michel}${}^{\calib}{}^{\caliw}$ 
   {and \bf Andrea Puhm}${}^\blacklozenge{}^\diamondsuit{}^\odot$
   }
  \bigskip
    \bigskip
        \centerline{\em${}^{\calib}$ Mani L. Bhaumik Institute for Theoretical Physics}
\centerline{\em Department of Physics and Astronomy, University of California}
\centerline{\em Los Angeles, CA 90095}
   \bigskip
   \centerline{\em${}^{\caliw}$ Department of Physics, University of California}
 \centerline{\em Santa Barbara, CA 93106}
   \bigskip
     \centerline{\em${}^\blacklozenge$       
     CPHT, \'{E}cole Polytechnique, CNRS}
     \centerline{\em 91128, Palaiseau, France
}
  \bigskip
 \centerline{\em${}^\diamondsuit$ Jefferson Physical Laboratory, Harvard University}
 \centerline{\em Cambridge, MA 02138}
 \bigskip
  \centerline{\em${}^\odot$ Black Hole Initiative, Harvard University}
  \centerline{\em Cambridge, MA 02138}

  \vskip2cm
  \end{center}

\begin{abstract}
  \noindent
Inspired by the recent work of Bao and Ooguri (BO), we study the distinguishability of the black hole microstates from the thermal state as captured by the average of their relative entropies: the Holevo information. Under the assumption that the vacuum conformal block dominates the entropy calculation, BO find that the average relative entropy vanishes on spatial regions smaller than half the size of the CFT. However, vacuum block dominance fails for some microstates of the $M=0$ BTZ black hole. We show that this renders the average relative entropy nonzero even on infinitesimal intervals at $O(c^0)$.
\end{abstract}

 \end{titlepage}

\setcounter{tocdepth}{2}

\tableofcontents

\section{Introduction}
Relative entropy is among the most powerful tools available for
characterizing the distinguishability of two quantum states, and has
recently found broad application in holography: its equality on bulk
and boundary convincingly identifies the entanglement wedge as the
dual to a given boundary region~\cite{Dong:2016eik}; its positivity
and monotonicity
leads to proofs of the average null energy condition~\cite{Faulkner:2016mzt} and the generalized second law~\cite{Wall:2010cj}; its
expansion about the vacuum leads to a derivation of Einstein's equations from
the dynamics of
entanglement~\cite{Lashkari:2013koa,Blanco:2013joa,Faulkner:2013ica,Hijano:2014sqa,Faulkner:2017tkh,Haehl:2017sot}. One might hope to use relative entropy to further probe questions related to the black hole information paradox, such as how easily the black hole microstates can be distinguished from the coarse-grained thermal state of the black hole.

Unfortunately, despite our ability to make general statements about
its properties, relative entropy is not an easy quantity to
compute in black hole microstates. Unlike its less-useful cousins -- the entanglement and Renyi
entropies -- it lacks a simple replica expression as a four-point function, and instead requires calculation of $n$-point correlators of heavy operators~\cite{Lashkari:2015dia,Sarosi:2016oks}.

Bao and Ooguri~\cite{Bao:2017guc} have recently identified an analog quantity which can be more easily computed -- the \emph{Holevo information}~$\chi$.  Given an ensemble $\rho=\sum_i p_i \rho_i$ of density matrices $\rho_i$, the Holevo information is the average entropy of the $\rho_i$ relative to $\rho$:
\begin{align}
\label{defchi}
\chi(\rho) &= \sum_i p_i S(\rho_i||\rho)\nonumber\\
&= S(\rho) - \sum_i p_i S(\rho_i)\,.
\end{align}
Here the $\rho_i$ and $\rho$ are normalized density matrices, the $p_i$ form a normalized probability distribution, and in the second line we used the definition of relative entropy:
\begin{equation}
S(\sigma||\rho) = -\tr(\sigma\log\rho) + \tr(\sigma\log\sigma).
\end{equation}
Since the average relative entropy $\chi$ is equal to the average difference in von Neumann entropy between $\rho$ and the $\rho_i$, computation of $\chi$ reduces to the computation of tractable (heavy-light) four-point functions.

Bao and Ooguri estimated this quantity for the thermal density matrix of a black hole in anti-de Sitter space at temperature $1/\beta$:
\begin{equation}
\rho_{\text{BH}}= Z^{-1} \sum_i e^{-\beta E_i} \ket{E_i}\bra{E_i}.
\label{eq:rhoBH}
\end{equation}
Here the microstates $\rho_i$ are the energy eigenstates
$\ket{E_i}\bra{E_i}$ weighted by a Boltzmann factor $p_i\propto
e^{-\beta E_i}$. They can be perfectly distinguished by an observer who has access to the entire dual CFT, but a less omniscient observer has a more difficult task.

Within a ball-shaped subregion $A_\ell$ of the CFT with diameter $\ell$, average distinguishability is characterized by the Holevo information of the reduced density matrix $\rho_{\ell}$ obtained by tracing over the complement of the region:
\begin{equation}
\rho_{\ell} \equiv \tr_{{\bar{A}_\ell}} \rho.
\label{eq:rhoBHMSinterval}
\end{equation}
In general dimension there is no known way to compute the $S(\rho_{\ell})$ in field theory and so
one must resort to holographic arguments, using the
HRRT~\cite{Ryu:2006bv,Hubeny:2007xt} formula in either the black hole
background or the (generically unknown) background created by a heavy
operator. However, in large-$c$ CFTs the
vacuum block contribution to the heavy-heavy-light-light four-point function
can be computed using the bootstrap \cite{Fitzpatrick:2014vua}, which in turn makes it possible to compute
$S(\rho_{i,\ell})$ in CFT~\cite{Asplund:2014coa} if one makes the assumption that
conformal blocks other than the vacuum do not contribute. Bao and Ooguri use this approximation to obtain~\cite{Bao:2017guc} 
\begin{equation}
 \chi(\rho_{\text{BH},\ell})\approx\left\{\begin{array}{ll} 0\, \quad & \ell \leq \pi R\,\\ \frac{c}{3} {\rm log} \left[\frac{\sinh(\pi \ell/\beta)}{\sinh\left(\pi (2\pi R-\ell)/\beta\right)}\right]\,
 \quad & \pi R<\ell <\ell_{crit}\,\\ S_{BH}\, \quad & \ell \geq \ell_{crit}\,,\end{array}\right.
 \label{eq:chiBO}
\end{equation}
where $c$ is the central charge of the CFT and $\ell_{crit}$ is a critical length that follows from the homology constraint in the Ryu-Takayanagi formula. We have reintroduced the radius $R$ of the spatial circle of the CFT and corrected several factors in this expression.

The vacuum block does not dominate at subleading orders in
$1/c$~\cite{Perlmutter:2015iya} and so one would naturally expect~\eqref{eq:chiBO} to receive
corrections. However, we will show that $1/c$ corrections to
\eqref{eq:chiBO} are also
induced by competition with the vacuum block in $S(\rho_{i,\ell})$
at {\it leading} order in certain states. This generates $1/c$
corrections to the average relative entropy once the rarity of these states is taken into account.

We will study these corrections in the $\beta\rightarrow
\infty$ limit corresponding to the $M=0$ BTZ black
hole~\cite{Banados:1992wn,Banados:1992gq}, which has a microscopic
description in terms of D-branes whose near-horizon geometry is
AdS$_3\times S^3$~\cite{Strominger:1996sh} (the $S^3$ is necessary in the microscopic formulation of the theory and will play an important role). We restrict to infinitesimal intervals
$\ell\rightarrow 0$ in the CFT, where the structure of entanglement
entropy can be made particularly
explicit~\cite{Calabrese:2010he,Giusto:2014aba}. The correction to $\chi$
that we estimate comes from states $\ket{i}$ whose R-charges $J$ scale with
$c$, corresponding to gravitational solutions with angular momentum on
the $S^3$. The expectation values $\langle J\rangle_i$ grow with
the central charge, violating the assumption of vacuum block dominance.

Our main result is an $O(c^0)$ correction to the average relative entropy in the small interval limit $\ell\rightarrow 0$:
\begin{equation}
\boxed{\chi(\rho_{\text{BH},\ell}) = \alpha \left(\frac{\ell}{R}\right)^2+\dots}
\end{equation}
where $\alpha$ is an $O(c^0)$ coefficient. The ellipses denote further
positive corrections: at $O(\ell^2)$ from BPS operators other than $J$
with conformal weights $h+\bar h=1$, and at higher orders in $\ell$ from higher-dimension operators. We also resolve the minor discrepancy between the holographic and field-theoretic computations of~\cite{Giusto:2014aba}.

The rest of this note is organized as follows. In \S~\ref{sec:Bao-Ooguri} we briefly review the work of Bao and Ooguri and their input assumptions. In \S~\ref{sec:fuzz} we review the work of Giusto and Russo on entanglement entropy in black hole microstates, which leads to our main result. We conclude in \S~\ref{sec:discussion} with some broader context. In the appendix we give two derivations of the density of states at generic R-charges: the first is intuitive and can be visualized; the second, which extends the thermodynamic analysis of Balasubramanian et al~\cite{Balasubramanian:2005qu}, allows us to systematically compute finite-$N$ corrections.

\section{Review of the Bao-Ooguri Estimate}\label{sec:Bao-Ooguri}
In the AdS$_3$/CFT$_2$ correspondence the BTZ black hole is dual to the thermal ensemble~\eqref{eq:rhoBH}. 
Following~\cite{Bao:2017guc}, we will be interested in characterizing the distinguishability\footnote{By distinguishability we mean that there exists an operator $O$ such that ${\rm tr}(\rho_i O)\neq {\rm tr}(\rho_\text{BH} O)$.} of the microstates $\rho_i$ from the ensemble~\eqref{eq:rhoBH} on an interval of length $\ell$ in the CFT, which we take to live on a circle of radius $R$. It is captured by the relative entropy
\begin{equation}
 S(\rho_{i,\ell}||\rho_{\text{BH},\ell})=-{\rm tr}(\rho_{i,\ell} {\rm log} \, \rho_{\text{BH},\ell})+{\rm tr}(\rho_{i,\ell}{\rm log}\, \rho_{i,\ell})\,,
\end{equation}
where $\rho_{i,\ell}$ and $\rho_{\text{BH},\ell}$ are the reduced density matrices defined in~\eqref{eq:rhoBHMSinterval}. The average of this quantity -- the Holevo information -- is equal to the average difference of von Neumann entropies:
\begin{align}
\label{eq:holevo}
\chi(\rho_{\text{BH},\ell}) &= \sum_i p_i S(\rho_{i,\ell}||\rho_{\text{BH},\ell})\nonumber\\
&= S(\rho_{\text{BH},\ell}) - \sum_i p_i S(\rho_{i,\ell})\,.
\end{align}
It was recently argued~\cite{Bao:2017guc} that the Holevo information
is zero on intervals smaller than half of the circle,
$\ell < \pi R$. We will argue in the next section that $1/c$
corrections make the Holevo information nonzero even on infinitesimal
intervals but first, we review the arguments
of~\cite{Bao:2017guc}, highlighting the assumption that will be violated in our scenario.

In order to evalute~\eqref{eq:holevo} one needs the von Neumann entropies of the reduced density matrices. When the CFT is holographic, these entropies are determined by the HRRT formula~\cite{Ryu:2006bv,Hubeny:2007xt},
\begin{equation}
\label{HRRT}
 S_{RT}(\rho_\ell)=\frac{{\rm length}(\Gamma_{\ell})}{4G}\,.
\end{equation}
Here $\Gamma_{\ell}$ is the extremal surface in the bulk dual of $\rho$, homologous to the boundary region $A_\ell$.

When $\rho=\rho_\text{BH}$ the dual geometry is the BTZ black hole. The homology constraint then implies a critical length $\ell_{crit}$ where the extremal surface splits into two disconnected pieces, one wrapping the black hole and another homologous to the complementary region:
\begin{equation}
 S(\rho_{\text{BH},\ell})=\left\{\begin{array}{ll}  S_{RT}(\rho_{\text{BH},\ell})&\quad  \ell<\ell_{crit}\\ S_{RT}(\rho_{\text{BH},2\pi R-\ell}) +S_{BH}& \quad \ell \geq \ell_{crit}.\end{array}\right.
 \label{eq:SRTensemble}
\end{equation}

The bulk dual of a generic microstate is not known, but one can still
study the $S(\rho_{i,\ell})$ using field-theoretic methods. Bao
and Ooguri~\cite{Bao:2017guc} estimated the microstate entropies using the results of
\cite{Asplund:2014coa}, which starts from the replica trick for entanglement entropy in 1+1d CFTs~\cite{Calabrese:2009qy}:
\begin{equation}
\label{eq:entropy}
 S(\rho_{i,\ell})=-{\rm tr} \, \rho_{i,\ell} {\rm log}\, \rho_{i,\ell}= \lim_{n\to 1} \frac{1}{1-n} {\rm log}\left[\mathcal{Z}^{-1}\langle \Psi_i(\infty) \T_n(u,\bar u) {\T}_{-n}(v,\bar v) \Psi_i(0) \rangle\,\right]\,.
\end{equation}
Here $u$ and $v$ are the endpoints of the interval, $\ket{\Psi_i}=\ket{i}^{\otimes n}$, the $\T_n$ are twist operators with weights $h_\T = \bar{h}_\T=\frac{c}{24}\frac{n^2-1}{n}$, and we have made explicit the
renormalization factor $\mathcal{Z}$.

The four-point function in~\eqref{eq:entropy} can be expanded in conformal blocks\footnote{The computation of~\eqref{eq:vNvac} in~\cite{Asplund:2014coa} starts on the plane and transforms to the cylinder. In the next section we will evaluate~\eqref{eq:entropy} directly in the theory on the cylinder.}~\cite{Asplund:2014coa}, using a scale transformation to move the operators to $(0,1,x,\infty)$:
\begin{equation}
\langle \Psi_i(\infty) \T_n(x,\bar x) {\T}_{-n}(1) \Psi_i(0) \rangle = \sum_{\mathcal{O}_p} a_{p}^{(i)} \, \mathcal{F}(nc;nh_i,h_\T,h_p;1-x) \bar{\mathcal{F}}(nc;n\bar h_i,\bar h_\T,\bar h_p;1-\bar x)\,.
\label{eq:confblock}
\end{equation}
Here the sum runs over Virasoro primaries, $a_{p}^{(i)}=C_{iip}C_{\T
  {\T}}^{p}$, and $x=u/v$ is the conformal cross-ratio.  We have chosen to expand in the $t$-channel to anticipate our eventual interest in the small-interval limit.

In large-$c$ conformal field theories this expression can be evaluated at
leading order in~$c$ under the additional assumption of vacuum block dominance~\cite{Hartman:2013mia}, i.e. that the sum is well-approximated by the contribution of the identity and its descendants:
\begin{equation}
\label{eq:vacblk}
\lim_{c\rightarrow\infty}\, \langle \Psi_i(\infty) \T_n(x,\bar x) {\T}_{-n}(1) \Psi_i(0) \rangle \stackrel{?}{\approx}\mathcal{F}(nc;nh_i,h_\T,0;1-x) \bar{\mathcal{F}}(nc;n\bar h_i,\bar h_\T,0;1-\bar{x})\,.
\end{equation}
This approximation is valid (in some range of $x$) only in states where $a_p^{(i)}$ grows slowly with $c$ for all light primaries $\O_p$~\cite{Hartman:2013mia,Asplund:2014coa}. In the next section we consider
a class of black hole microstates that violate this condition and discuss
their impact on $\chi$.

If one simply assumes vacuum block dominance, the correlator
in~\eqref{eq:entropy} can be computed in the limit $n\rightarrow 1$~\cite{Asplund:2014coa} using the known form of the heavy-heavy-light-light vacuum block~\cite{Fitzpatrick:2014vua} in the channel in which it dominates. The result is the vacuum block contribution to $S(\rho_{i,\ell})$:
\begin{equation}
 S(\rho_{i,\ell})|_\text{vac}=\frac{c}{3} {\rm log} \left[\frac{\beta_i}{\pi\,  \epsilon} \sinh\left(\frac{\pi \, {\rm min}(\ell,2\pi R-\ell)}{\beta_i}\right)\right]\,,
 \label{eq:vNvac}
\end{equation}
where
\begin{equation}
 \beta_i=\frac{2\pi}{\sqrt{24h_i/c-1}}
\end{equation}
and $\epsilon$ is the UV cutoff. When $\ell>\pi R$ the vacuum block dominates in the channel obtained by replacing $\ell$ with $2\pi R-\ell$~\cite{Asplund:2014coa}, leading to the min in~\eqref{eq:vNvac}. This is the expression that is used in
\cite{Bao:2017guc} to obtain the estimate~\eqref{eq:chiBO},
with the weight $h_i$ of the microstate fixed by demanding $\beta_i =
\beta$. It agrees precisely with the holographic entropy~\eqref{HRRT} in
the BTZ geometry if one relaxes the homology constraint~\cite{Asplund:2014coa}.

Collecting~\eqref{eq:SRTensemble} and~\eqref{eq:vNvac}, the vacuum block contribution to $\chi$ is
\begin{equation}
 \chi(\rho_{\text{BH},\ell})|_\text{vac}=\left\{\begin{array}{ll} 0 & \quad
                                                          \ell<\pi R\\
                                      S_{RT}(\rho_{\text{BH},\ell})-S_{RT}(\rho_{\text{BH},2\pi R-\ell})=\frac{c}{3}
                                      {\rm log} \left[\frac{\sinh(\pi
                                      \ell/\beta)}{\sinh\left(\pi(2\pi R-\ell)/\beta\right)}\right]
                                                        & \quad \pi R\leq \ell \leq \ell_{crit}\\ S_{BH} & \quad \ell > \ell_{crit}.\end{array}\right.
 \label{eq:HolevoBO}
\end{equation}
We will see that the failure of vacuum block
dominance makes $\chi\sim O(c^0)$ even when $\ell\ll \pi R$.

\section{Fuzziness and Distinguishability}
\label{sec:fuzz}

\subsection{Entanglement entropy in the $M=0$ BTZ microstates}
\label{ssec:EE}

We will study the Holevo information in a string theory setup where the microstates of the black hole can be identified explicitly: the near-horizon limit of the D1-D5 brane system, whose CFT description includes 1/4-BPS states that are
dual to microstates of the $M=0$ BTZ black hole ($\times\, S^3 \times\,M_4$). The study of these states and their identification with dual geometries~\cite{Lunin:2001jy} is a rich subject with a long and tempestuous
history;\footnote{See~\cite{Mathur:2005zp,Bena:2007kg,Mathur:2008nj,Balasubramanian:2008da,Bena:2013dka}
  for reviews
  and~\cite{Chen:2014loa,Eperon:2016cdd,Marolf:2016nwu,Martinec:2017ztd}
  for some recent developments.} we will not make use of this
identification here. Our results follow solely from the properties of 1/4-BPS operators in the CFT\footnote{The operators we study are chiral primaries~\cite{Lerche:1989uy}.} and are independent of (but consistent with) the identification of certain BPS operators with bulk geometries. For more details of the CFT see e.g.~\cite{Lunin:2001pw,Lunin:2001jy,David:2002wn,Balasubramanian:2005qu,Kanitscheider:2007wq} and references therein. 

The field theory description of the microstates is obtained by
considering type IIB string theory on $S^1\times M_4$ with $N_1$ D1s wrapping the $S^1$ and $N_5$ D5s wrapping all the compactified directions, with the closed string sector decoupled by the near-horizon limit. Taking $M_4$ much smaller than the
$S^1$, the theory becomes a 1+1d SCFT with $\mathcal{N}=(4,4)$ SUSY,
whose central charge $c=6N_1 N_5\equiv 6N$ is determined by the number
of D1 and D5 branes. At weak coupling the SCFT is conjectured to
be a nonlinear sigma model on the symmetric product orbifold $(M_4)^{N}/S_{N}$
\cite{Dijkgraaf:1998gf}; at strong coupling it is dual to supergravity in
AdS$_3\times\, S^3 \times\,M_4$~\cite{Maldacena:1997re}.

The microstates of the $M=0$ BTZ black hole correspond to the Ramond
ground states of this D1-D5 CFT~\cite{Strominger:1997eq,Lunin:2002iz}. Supersymmetry protects their dimensions and $SU(2)_L\times SU(2)_R$ R-charges, which can therefore
be extrapolated from the orbifold theory to the region of moduli space
where the theory has a gravitational dual. The R-charges of the black hole microstates are bounded by $-\frac{N}{2}<
J^3_L, J^3_R < \frac{N}{2}$~\cite{Lerche:1989uy} and correspond to
angular momenta on the $S^3$ in the bulk~\cite{Lunin:2001fv}. Most states
have $ J^3_{L/R} \gtrsim O(\sqrt{N})$, violating vacuum block dominance.

As in any CFT corrections to the vacuum entropy are particularly easy to isolate in the small-interval limit, where the physics is determined by the twist operator OPE~\cite{Calabrese:2010he}. Since $\T_n \T_{-n}$ has vanishing twist, only untwisted operators\footnote{There are orbifolds and then there are orbifolds. Here we are in the theory on $\left((M_4)^N/S_N\right)^n/\mathbb{Z}_n$.} $\O=\O_{p_1}^{(1)}\otimes\dots\otimes \O_{p_n}^{(n)}$ appear:
\begin{eqnarray}\label{eq:4.5GR}
\T_n(w,\bar w) \T_{-n}(w',\bar w')&=&\frac{1}{|w-w'|^{4h_\T}} \Big[1+\sum_p\sum_{r=1}^n (w-w')^{h_p} (\bar w-\bar w')^{\bar h_p} C_{\T \T}^{p_r} \, \O_{p}^{(r)}\\
&&+ \sum\limits_{\substack{p,q\\r \neq s}} (w-w')^{h_p+h_{q}}(\bar w-\bar w')^{\bar h_p+\bar h_{q}} C_{\T \T }^{(p_r,q_s)} \, \O_{p}^{(r)}\otimes \O_{q}^{(s)} +\dots\Big],\nn
\end{eqnarray}
with further terms subleading as $w\rightarrow w'$. Here $r,s$ are replica indices and the expansion has been organized according to how many of the tensor factors in $\O$ are trivial (e.g. on the second line the operator is $\O_p$ on the $r^\text{th}$ factor tensored with $\O_{q}$ on the ${s}^\text{th}$ factor, tensored with the identity on the remaining factors). This expansion was studied by~\cite{Giusto:2014aba} in the Ramond ground states dual to the microstates of the $M=0$ BTZ black hole; we review their treatment, amending two points.

In any state $\ket{i}$, the entanglement entropy of a small interval can be computed by plugging~\eqref{eq:4.5GR} into~\eqref{eq:entropy}. We use coordinates $w$ on the cylinder related to the plane coordinates $z$ via $z=e^{iw}$, with the parametrization

\be
w= \frac{\phi}{R} + i\tau,\quad\quad \tau\in \R,\quad\quad \phi \equiv \phi+2\pi R.
\ee
The operators in~\eqref{eq:4.5GR} are inserted at $\phi-\phi'=\ell$ on a constant-$\tau$ slice.

As $\ell\rightarrow 0$ the dominant contributions to~\eqref{eq:4.5GR} come from the lightest operators in the theory. The leading correction naively appears to be the first non-vacuum term in
\eqref{eq:4.5GR}, but
this term vanishes for all primaries since $C_{\T \T
  p_r}=0$ if $\O_p$ is primary.\footnote{This follows from the vanishing of primary one-point functions on the plane. Generically one must keep track
  of descendant contributions to this term, and the stress tensor will make an
  $O(\ell^2)$ contribution to the entanglement entropy. However, $\langle
  T\rangle = 0$ in the Ramond ground states that we study in this paper. All other
  descendants in the D1-D5 CFT will contribute to this term at higher
  orders in~$\ell$.} In the $M=0$ BTZ microstates, the leading correction to the entropy
comes from a sum over the lightest untwisted primaries with two tensor factors nontrivial. When these operators have nontrivial expectation values they will contribute to the entanglement entropy. The OPE coefficients $C_{\T \T}^{(p_r,q_s)}$ can be determined by computing the expectation value of $\O_{p}^{(r)}\otimes \O_{q}^{(s)}$ on the replica manifold via a conformal map~\cite{Calabrese:2010he}, then using the inverse of the Zamolodchikov metric $\g_{ab}=\langle O_a|O_b\rangle$ to raise the index~\cite{Giusto:2014aba}.

We would like to compute the microstate entropies $S(\rho_{i,\ell})$ in
the gravity regime, but the orbifold CFT is at zero coupling.
Fortunately, there is strong evidence that the gravity computation of~\eqref{eq:4.5GR} can be extracted from the calculation in the orbifold theory. The idea is as follows~\cite{Giusto:2014aba}: most
operators acquire $O(c)$ anomalous dimensions at strong coupling and so will contribute
negligibly to~\eqref{eq:4.5GR}. However,
1/4-BPS operators are protected by supersymmetry: their dimensions and three-point functions are independent of the coupling\cite{Baggio:2012rr}. To extrapolate results from the orbifold theory to gravity we
therefore keep only the contributions from these BPS operators (we
will leave the 1/4 implicit for brevity). It was shown in~\cite{Giusto:2014aba} that this prescription agrees exactly with the results obtained from a generalization of the HRRT formula, apart from the vacuum piece we will correct.

The lightest BPS operators in the D1-D5 CFT have $h+\bar h=1$ and thus appear at $O(\ell^2)$ in the twist OPE~\eqref{eq:4.5GR}. Inserting~\eqref{eq:4.5GR} into the formula for the entropy~\eqref{eq:entropy} and evaluating the sum over $r$ and $s$ in this sector exactly as in~\cite{Giusto:2014aba}, at $O(\ell^2)$ we find\footnote{Note that we have obtained a slightly different expression from the corresponding equations (4.9-10) of~\cite{Giusto:2014aba}, which would have $\langle \O_{p'}\rangle_i \langle \O_{q'}\rangle_i$ inside the sum in our~\eqref{mpq} instead of $\langle \O_{p}\rangle_i \langle \O_{q}\rangle_i$ outside. This does not affect their results since $\mathfrak{g}_{ab}$ is block-diagonal in the space of BPS operators with $h+\bar h =1$.}
\begin{equation}
\bra{\Psi_i} \T_n(w,\bar w) \T_{-n}(w',\bar w') \ket{\Psi_i} =\left(\frac{R}{\ell}\right)^{\frac{c}{6}\frac{n^2-1}{n}} \left[1+\sum_{p,q} \left(\frac{\ell}{R}\right)^2\,  \frac{(n^2-1)}{24n} \, \mathcal{M}_{(p,q)}\, \langle\O_p \rangle_i \langle \O_{q} \rangle_i+\dots \right]
 \label{eq:fuzz4ptdim1}
\end{equation}
where $\mathcal{M}$ is the contraction of the Zamolodchikov metric with the plane two-point function:
\begin{equation}
\label{mpq}
\mathcal{M}_{(p,q)} = \g^{\{(p,q),(p',q')\}} \bra{0} \O_{p'}(1) \O_{q'}(0)\ket{0}_\text{plane}.
\end{equation}
Among the lightest BPS operators are the holomorphic and antiholomorphic
$SU(2)_L\times SU(2)_R$ R-symmetry currents $J^{\alpha}_L$ and
$J^{\alpha}_R$. For simplicity we will focus on corrections to the Holevo information
coming from the $\alpha=3$ components alone, neglecting the rest of the lightest BPS operators; we will abbreviate
$ J^3_{L/R}\equiv J_{L/R}$. Our computation of these
corrections will be enough to show that $\chi\gtrsim O(c^0)$ at
small $\ell$, since the other operators that contribute at $O(\ell^2)$ make manifestly positive
contributions to $\chi$.

To obtain $S(\rho_{i,\ell})$ from~\eqref{eq:fuzz4ptdim1} we must include the renormalization factor $\mathcal{Z}$, which was omitted in~\cite{Giusto:2014aba}. In the theory on the plane, the renormalization factor $\mathcal{Z}_\text{plane} = \epsilon^{-4h_\T}$ is introduced in order to make the entropy dimensionless~\cite{Calabrese:2009qy}. However, there is no need to fix the units in~\eqref{eq:fuzz4ptdim1} and so it seems we are free to choose $\mathcal{Z}_\text{cyl}$ as we please. We will fix it here by demanding that the leading divergence in $S(\rho_{i,\ell})$ approaches the vacuum entropy on the plane when $\ell\ll R$. This scheme is natural in light of the conformal symmetry: we should get the same results regardless of whether we do the computation directly in the theory on the cylinder, or by transforming the four-point function on the plane normalized by $\mathcal{Z}_\text{plane}$. Thus we take

\be
\label{zcyl}
\mathcal{Z}_\text{cyl} = \left(\frac{\epsilon}{R}\right)^{-4h_\T}.
\ee

Taking $\mathcal{Z}_\text{cyl}$ into account, focusing just on the contribution of $J_{L/R}$ and following the computation of $\mathcal{M}_{(J,J)}=12/c$ in~\cite{Giusto:2014aba}, we find
\begin{eqnarray}
\label{jent}
S(\rho_{i,\ell})&=&\frac{c}{3} {\rm log}\frac{\ell}{\epsilon} - \frac{1}{c}\left(\frac{\ell}{R}\right)^2 \left(\langle J_L\rangle_{i}^2+\langle  J_R\rangle_{i}^2\right)-\dots
\end{eqnarray}
where $\dots$ denotes the contribution of the other BPS and higher-dimension operators. We have found a different vacuum contribution than~\cite{Giusto:2014aba} due to our inclusion of the renormalization factor~\eqref{zcyl}; with this minor correction, the result agrees exactly with the holographic computation in~\cite{Giusto:2014aba} of the
entropy at $O(\ell^2)$ using the deformed HRRT surface
in the gravity background dual to the microstate. We
will not however make use of this interesting fact.

Reintroducing the contributions of the remaining BPS operators with $h + \bar h = 1$,
\begin{eqnarray}
S(\rho_{i,\ell})
&=& S(\rho_{i,\ell})|_{\rm vac} + 
    S(\rho_{i,\ell})|_\text{BPS}+\dots\,.
\label{eq:Svacchiral}
\end{eqnarray}
This is the exact microstate entropy at $O(\ell^2)$.

\subsection{Average relative entropy of the $M=0$ BTZ microstates}
\label{ssec:holevo}

The microstates $\ket{J_L,J_R}$ of the D1-D5 black hole are labelled by their left and right R-charges $(J_L, J_R)$. We compute the average relative entropy $\chi$ of the zero-temperature density matrix
\begin{equation}
 \rho_\text{BH}=\sum_{J_L,J_R}p\left(J_L,J_R\right)\ket{J_L,J_R}\bra{J_L,J_R} \equiv \sum_{J_L,J_R}p\left(J_L,J_R\right)\ \rho_{(J_L,J_R)}
 \label{eq:rhoBHJtJ}
\end{equation}
dual to the $M=0$ BTZ black hole $\times\ S^3$~\cite{Balasubramanian:2005qu}. We have left the other quantum numbers implicit as they do not affect the contribution we compute. The probability $p\left(J_LJ_R\right)$ that a microstate has charges $(J_L,J_R)$ is
the number of states $d_{J_L,J_R}$ with those charges divided by the total number of states, i.e.
\begin{equation}
\label{eq:pd}
p\left(J_L,J_R\right) \propto d_{J_L,J_R}.
\end{equation}

We will compute contributions to the average relative entropy of $\rho_\text{BH}$ restricted to an interval of length $\ell$,
\begin{eqnarray}
 \chi(\rho_{\text{BH},\ell})&=& \sum_{J_L,J_R}p\left(J_L,J_R\right)\ S(\rho_{(J_L,J_R),\ell}||\rho_{\text{BH},\ell})\nonumber\\ &=&S(\rho_{\text{BH},\ell}) - \sum_{J_L,J_R}p\left(J_L,J_R\right)\ S(\rho_{(J_L,J_R),\ell})\,.
\end{eqnarray}
In the limit $\ell\rightarrow 0$ this inherits the spectral structure of~\eqref{eq:Svacchiral}:
\begin{equation}
 \chi(\rho_{\text{BH},\ell})=
 \chi(\rho_{\text{BH},\ell})|_\text{vac}+\chi(\rho_{\text{BH},\ell})|_\text{BPS}+\dots
\end{equation}
where $\dots$ denotes subleading terms in $\ell$. The first term
vanishes as $\ell\rightarrow 0$~\cite{Bao:2017guc} because the vacuum block contribution is equal to the thermal entropy when $\ell<\pi R$. The second is $O(\ell^2)$:
\begin{equation}
\label{eq:chi}
  \chi(\rho_{\text{BH},\ell})|_\text{BPS}=\frac{1}{c} \left(\frac{\ell}{R}\right)^2
  \sum_{J_L,J_R} p\left(J_L,J_R\right) \left( J_L^2+
    J_R^2  \right)+\dots\,.
\end{equation}
We will ignore the (positive) contributions of the other 
BPS operators with $h+\bar h=1$ and so obtain only a lower bound.

\begin{samepage}
Specifying to $M_4=T^4$ for concreteness, in the appendix we find the density of states\footnote{The density of states had previously
  been computed~\cite{Balasubramanian:2005qu} in the regime where only one linear combinations of the
  R-charges, $J_+\equiv J_L + J_R$, is $O(N)$:
\begin{equation}
\label{footnotedos1j}
d_{J_L,J_R}= e^{2\pi \sqrt{2}\sqrt{N-|J_+|}}
\end{equation}
at leading order in $N$. Our more general expression~\eqref{eq:pjtj} implies a logarithmic correction to the density of states at fixed $J_+\sim O(N)$ from the sum over the other
linear combination $J_- \equiv J_L - J_R$:
\begin{equation}
\int_{-N}^{N} d J_-\ d_{J_L, J_R} \propto
e^{2\pi\sqrt{2}\sqrt{N-|J_+|}}\sqrt{N-|J_+|}.
\end{equation}
\label{jfootnote}
} 
\begin{equation}
\label{eq:pjtj}
d_{J_L,J_R}=  e^{2\pi\sqrt{2} \sqrt{N-|J_L+J_R|-|J_L-J_R|}}
\end{equation}
at leading order in $N$ when $J_{L/R}\sim O(N)$ (recall that $c=6N$).\end{samepage}

We can combine~\eqref{eq:pjtj} with~\eqref{jent} and~\eqref{eq:pd} to evaluate~\eqref{eq:chi}. Approximating the sum over
R-charges by an integral, we obtain
\begin{align}
 \chi(\rho_{\text{BH},\ell})|_\text{BPS}&\approx
 \frac{1}{c}\left(\frac{\ell}{R}\right)^2 \int_{-N/2}^{+N/2} \ dJ_L\ 
 dJ_R \ \ p\left(J_L,J_R\right) \, \left(J_L^2+ J_R^2\right) + \dots\nn\\
&= \alpha  \left(\frac{\ell}{R}\right)^2+\dots
\label{eq:chijtj}
\end{align}
where $\alpha\sim O(c^0)$ and $\dots$ denotes the contribution of the
other BPS operators with $h + \bar h = 1$. This is our main result.

There is an important caveat that does not affect our
conclusion. The density of states~\eqref{eq:pjtj} is only valid at
$J_{L/R}\sim O(N)$ while typical states of the D1-D5 CFT have $J_{L/R} \sim O(\sqrt{N})$~\cite{Balasubramanian:2005qu}.\footnote{There are also states with
  $J_{L/R} < O(\sqrt{N})$ but in such states the vacuum block
  dominates over $J$-exchange, so they make negligible
  contributions to the part of $\chi$ that we are estimating.}
In appendix~\ref{appendix:b} we show that when  $J_{L/R} \sim
O(\sqrt{N})$ the density of states is
\begin{equation}
d_{J_L,J_R}=  e^{2\pi\sqrt{2}\sqrt{N-\gamma_+ |J_L+J_R|-\gamma_- |J_L-J_R|}}\,, \qquad  0 < \gamma_{\pm} < 1
\end{equation}
where the $\gamma_{\pm}$ must be determined numerically but are smooth functions of $J_{L/R}$. We also show that if we use
this more accurate density of states instead of~\eqref{eq:pjtj} in the appropriate
region of the integral in~\eqref{eq:chijtj}, we still find that $\alpha\sim O(c^0)$.

\section{Discussion}\label{sec:discussion}

At the outset we motivated our investigation of the Holevo information by a hope that it would aid in study of the information paradox. Let us outline some steps in that direction.

Ultimately, one would like to calculate some bulk probe of the
distinguishability between the black hole microstates and the naive
black hole geometry dual to the thermal state. Relative entropy is
such a probe, but it is hard to calculate. However, we have seen that the
{\it average} relative entropy -- the Holevo information -- can be estimated quite
easily under sufficiently controlled circumstances. Fortunately, it is
a decent substitute for the relative entropy in addressing certain interesting questions. Suppose we are wondering whether or not most states of a particular black hole have a firewall near the horizon.
In this case one can simply ask whether $\chi$ is large in a region
that probes the horizon. Unfortunately, we do not know much about the microstates of black holes that may have firewalls,
and our approach has restricted us to both supersymmetry and small $\ell$.

One step towards a more realistic black hole is the $M>0$ BTZ
black hole dual to the \mbox{D1-D5-$p$} system. Its decoupling limit is the
same CFT we have studied but the microstates have nonzero momentum $N_p$ along the common D1-D5 direction. There are additional light BPS operators with nontrivial expectation values in these states, and since $\langle T\rangle\sim N_p\neq 0$ the stress tensor also contributes to $\chi$ at $O(\ell^2)$~\cite{Giusto:2015dfa}. Despite much recent progress (see e.g.~\cite{Bena:2014qxa,Martinec:2015pfa,Bena:2015bea,Bena:2016ypk,Bena:2017geu,Bena:2017upb,Tyukov:2017uig,Bena:2017xbt}) knowledge of the D1-D5-$p$ microstates is not comprehensive; one would still need to know how the expectation values are distributed in order to compute contributions to $\chi$. At the end of the day one would like to break supersymmetry and study evaporating black holes,
but knowledge of their microstates is even more sparse~\cite{Jejjala:2005yu,Bena:2011fc,Bena:2012zi,Bena:2016dbw,Bossard:2017vii}.

There are also prospects in higher dimensions. In AdS$_5$ there is a close analog of the D1-D5 system, the LLM geometries~\cite{Lin:2004nb}, which are dual to 1/2-BPS states of $\mathcal{N}=4$ super-Yang Mills. The analog of the black hole for these geometries is the superstar solution~\cite{Myers:2001aq,Balasubramanian:2005kk,Balasubramanian:2005mg}. While the LLM geometries are explicitly known, computation of the entanglement entropies seems to be difficult,\footnote{We thank Michael Gutperle, Mukund Rangamani and Joan Sim\'{o}n for discussion on this point.}
though there has been some recent progress
in a different duality frame~\cite{Kim:2016dzw}.

What about the vacuum block approximation? It is crucial to much
of the recent
technology
that has been developed to study both correlation functions and
bulk reconstruction in AdS$_3$~\cite{Fitzpatrick:2015dlt,Fitzpatrick:2016thx,Fitzpatrick:2016ive,Fitzpatrick:2016mtp,Anand:2017dav,Chen:2017dnl,Faulkner:2017hll}. We do not know whether or not this
affects the ability of these methods to make predictions about what
happens near black hole horizons -- one might suspect that the gravitational physics is sufficiently universal -- but it could be useful to extend their approach to WZW models. With such techniques one can probe much further into the bulk than we
have been able to here.

It is also natural to ask if $\chi$ itself has some interpretation in terms of ordinary (non-replica) correlation functions. While the relative entropy can be expressed in terms of the expectation value of the modular hamiltonian as $\Delta\langle K \rangle -\Delta S$, the first term drops out of the average. From the fact that two density matrices have a nonzero relative entropy we can conclude that there is some operator such that $\tr(\rho_i O)\neq \tr(\rho O)$, which must even be local since we consider small $\ell$, but the average expectation values are necessarily equal: $\sum p_i \tr(\rho_i O) = \tr(\rho O)$.

Last we discuss corrections to our result. First, $\chi$ receives
further positive contributions at $O(\ell^2)$ from other BPS operators with $h +
\bar h = 1$. It seems reasonable to conjecture that the
contributions of these operators do not increase $\alpha$ beyond $O(c^0)$, though they
might. One would need to know the distribution of their expectation values.

The second source of correction is more interesting. Our $O(c^0)$ result comes at the order typically associated with quantum effects in the bulk. For instance, at $O(c^0)$ one must account for the entanglement of bulk fields, which generically contains state-dependent pieces as in~\eqref{eq:Svacchiral} and so $\chi$ will be further corrected. Of course this is just the usual story of quantum corrections in the bulk  -- the same effect gives rise to mutual information at $O(c^0)$ between distant boundary intervals -- but it is intriguing that there is an interplay here with the sum over geometries.

\section*{Acknowledgements}
We are grateful for discussions with Ning Bao, Stefano Giusto, Tim Hsieh,
Don Marolf, Rodolfo Russo, Masaki Shigemori, Tomonori Ugajin, and especially Per Kraus and Mark Srednicki. BM was supported by NSF Grant PHY13-16748 and the Mani L. Bhaumik Institute at UCLA. AP acknowledges support from the Black Hole Initiative at Harvard University, which is funded by a grant from the John Templeton Foundation, and support from the Simons Investigator Award 291811 and DOE grant DE-SC0007870. AP also acknowledges the hospitality of the Weizmann Insitute of Science during ``Post Strings 2017", the Erwin Schr\"odinger Insitute during ``Quantum Physics and Gravity" and the Yukawa Institute of Theoretical Physics at Kyoto University during ``Recent Developments in Microstructures of Black Holes''.

\appendix

\section{Density of states at generic R-charges}

\subsection{Intuitive derivation of the density of states}

Supersymmetry protects the ground states of the D1-D5 CFT dual to the microstates of the $M=0$ BTZ black hole. This is relevant for our purposes because it implies that the number of states with R-charges $J_{L/R}$ remains constant as we go from the strongly-coupled regime, where the system is dual to gravity in AdS$_3\times\, S^3\times\, M_4$, to zero coupling, where the system is described by an orbifold theory on $(M_4)^N/S_N$. We can therefore use the orbifold description to obtain a particularly simple derivation of the density of states. We will need just one fact~\cite{Lunin:2001jy} about the ground states $\ket{\psi}$ in the orbifold theory: they are completely characterized by a set of integers,
\begin{equation}
\label{dospartition}
\ket{\psi\left(\left\{N_{n,i}\right\}\right)} \longleftrightarrow \prod_{n,i} \left( \Phi_n^{(i)}\right)^{N_{n,i}}.
\end{equation}
Here $\Phi_n^{(i)}$ is a twist operator permuting $n$ out of the $N$ copies of the seed theory (which is an SCFT on $M_4$) and $i$ runs over all the fields in the seed. Note that $n$ here has no connection to the replica index used in the entropy calculations in the main text.

\begin{figure}[ht!]
\begin{center}
 \includegraphics[width=0.4\textwidth]{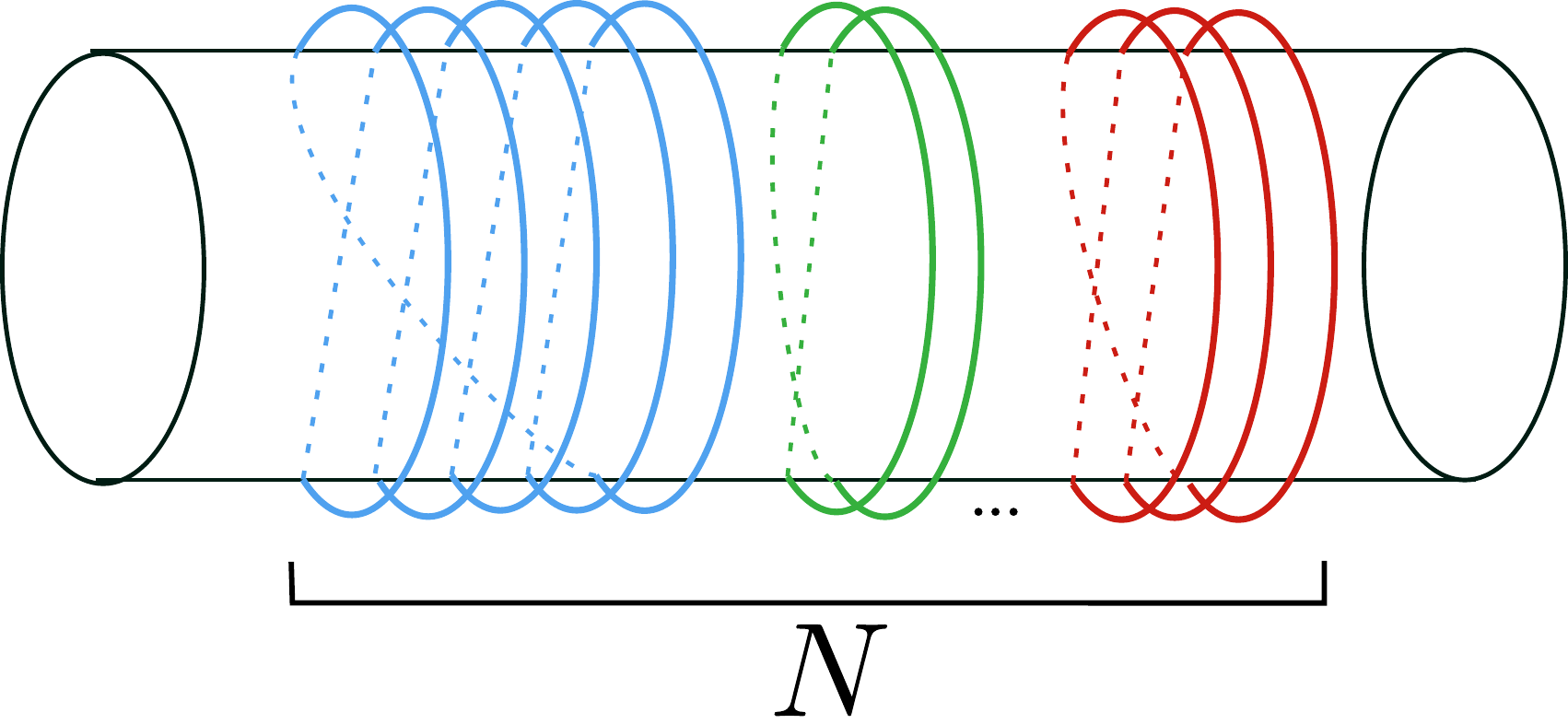} 
 \end{center}
 \caption{The black hole microstates correspond to colored partitions of $N$, which can be illustrated as a collection of multiwound colored strings: $N_{n,i}$ is the number of strings with winding number $n$ and color $i$.}
\label{fig:multistringJeq0}
\end{figure}

The integers $N_{n,i}$ can be chosen freely up to the constraint
\begin{equation}
\label{constraint}
\sum_{n,i} n N_{n,i} = N,
\end{equation}
which implies that the ground states $\ket{\psi}$ are in one-to-one correspondence with colored partitions of $N$, with $i$ labelling the different colors. This is depicted in figure~\ref{fig:multistringJeq0}.\footnote{The figures in this appendix can be read quite literally: the $\Phi^{(i)}_n$ map to the usual string creation operators $\alpha^{i}_{-n}$ under U-duality~\cite{Lunin:2001jy}.}

An asymptotic estimate of the number of uncolored partitions was derived by Hardy and Ramanujan~\cite{Hardy1918}:
\begin{equation}
P(N) = e^{\pi\sqrt{\frac{2N}{3}}}
\end{equation}
to leading order in $N$. We have one color for each boson and half a color for each fermion (due to bosonization), so the total number of colors is the central charge of the seed theory,
\be
N_B + \frac{N_F}{2} \equiv c_s.
\ee
The total number of microstates is then the number of partitions with $c_s$ colors:
\be
\label{dn}
P_{c_s}(N) = e^{2\pi\sqrt{\frac{c_s}{6}N}}
\ee
at large $N$. Notably, the logarithm of this quantity is the Bekenstein-Hawking entropy of the $M=0$ BTZ black hole~\cite{Dabholkar:2004yr} (once $\alpha'$ corrections have been taken into account).

\begin{figure}[ht!]
\begin{center}
 \includegraphics[width=0.5\textwidth]{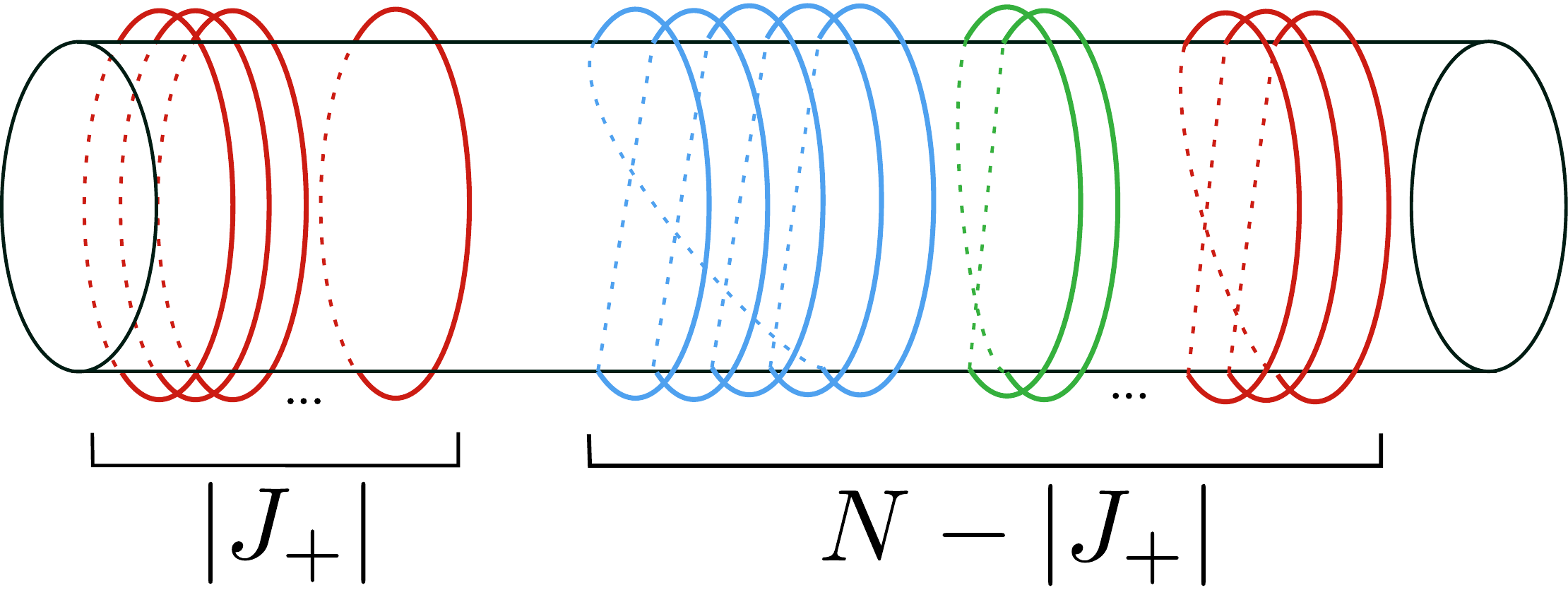} 
 \end{center}
 \caption{A typical state with positive $J_+ \sim O(N)$ and $J_-\sim 0$. The red strings carry $(J_+, J_-)$ charges $(1, 0)$ while the other strings carry other R-charges, including zero. The freedom to pick any colored partition of the remaining $N-|J_+|$ elements gives rise to the entropy \eqref{footnotedos1j}.}
\label{fig:multistringJp}
\end{figure}

One can extend this reasoning to obtain~\eqref{eq:pjtj}.\footnote{The argument given in this section was originally used in \cite{Balasubramanian:2005qu} to intuitively explain the density of states~\eqref{footnotedos1j}.} First, note that the fields $i$ are labeled by their R-symmetry charges. Focus on the four R-charged bosons, which carry
\begin{equation}
(J_L, J_R) = (\pm 1/2, \pm 1/2),\ (\pm 1/2, \mp 1/2).
\end{equation}
In terms of the linear combinations $J_\pm = J_L \pm J_R$, they carry
\be
(J_+, J_-) = (\pm 1,0),\ (0,\pm 1).
\ee
Suppose we want to estimate the number of states with fixed $J_+\sim O(N)$. This is a small fraction of the total number of states~\eqref{dn}, since a typical colored partition will not carry much R-charge; we must fix some of the coloring so that the corresponding state carries $J_+$.\footnote{The unfixed part will be R-neutral on average since the fields come in pairs with equal and opposite charges.} At large $N$, the dominant contribution comes from the configurations depicted in figure~\ref{fig:multistringJp}. These configurations have $|J_+|$ singly-wound strings whose color corresponds to either $(J_+, J_-) = (1,0)$ or $(-1,0)$, depending on the sign of $J_+$; the remaining $N_1 = N-|J_+|$ elements can be in an arbitrary configuration. The number of these configurations is the number of colored partitions of the unfixed remainder,
\be
P_{c_s}(N_1) = e^{2\pi\sqrt{\frac{c_s}{6}(N-|J_+|)}}
\ee
at large $N$. Configurations with some of the $J_+$ carried in multi-wound strings are subleading, since fewer elements will be left unfixed. Likewise subleading are configurations with $J_+$ carried by fermions, since the Pauli exclusion principle makes it impossible to carry all of $J_+$ in singly-wound fermionic strings. The collection of strings carrying $J_+$ can be thought of as a Bose-Einstein condensate~\cite{Balasubramanian:2005qu}.

\begin{figure}[ht!]
\begin{center}
 \includegraphics[width=0.75\textwidth]{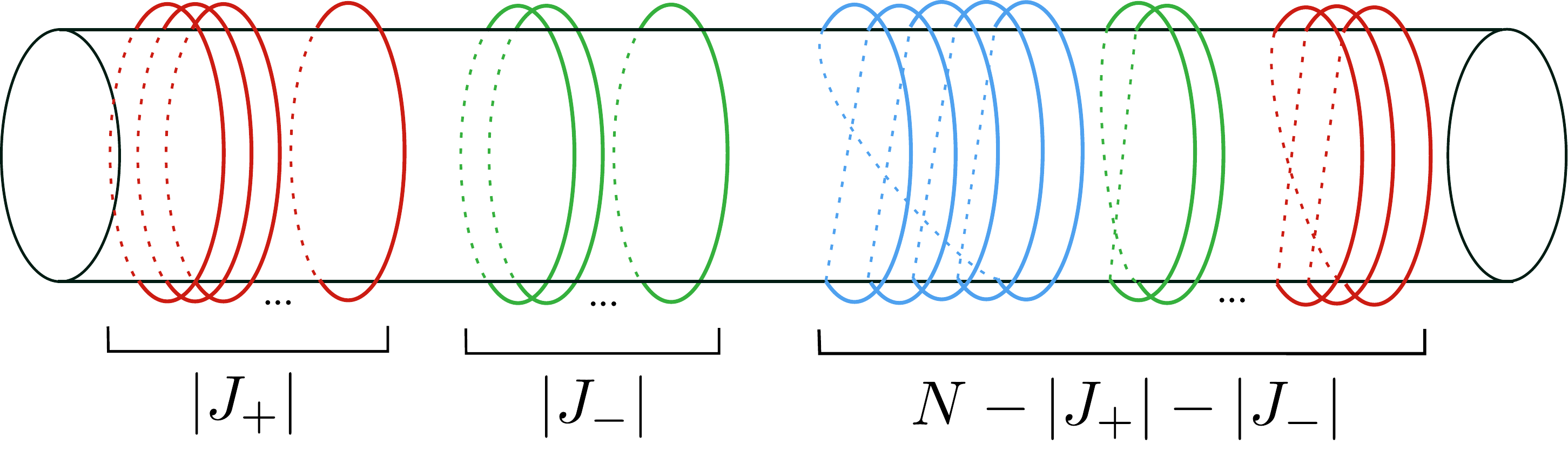} 
 \end{center}
 \caption{A typical state with positive $J_{\pm} \sim O(N)$. The red strings carry $(J_+,J_-)=(1, 0)$ and the green strings carry $(J_+,J_-)=(0,1)$. The freedom to pick any colored partition of the remaining $N-|J_+|-|J_-|$ elements gives rise to the entropy \eqref{eq:pjtj}.}
 \label{fig:multistringJpJm}
\end{figure}

To estimate the density of states at generic $J_{L/R}$ we must estimate the number of colored partitions with both $J_+$ and $J_-$ fixed. Following the same logic as above, most of these partitions will have the structure depicted in figure~\ref{fig:multistringJpJm}, with two condensates: one made of $|J_+|$ singly-wound strings carrying $(J_+,J_-) = (\pm 1,0)$, and another made of $|J_-|$ singly-wound strings carrying $(J_+, J_-) = (0,\pm 1)$, with the signs again determined by those of $J_\pm$. The entropy then comes from colored partitions of the remaining $N_2= N - |J_+| - |J_-|$ elements:
\be
d_{J_L,J_R} \approx P_{c_s}(N_2) = e^{2\pi\sqrt{\frac{c_s}{6}(N-|J_L+J_R| - |J_L - J_R|)}}.
\ee
The theory on $M_4=T^4$ has $c_s = 12$, which yields~\eqref{eq:pjtj}.

\subsection{Thermodynamic derivation of the density of states}\label{appendix:b}

Eq.~\eqref{eq:pjtj} can be confirmed by a direct calculation of the density of states, which also enables us to study corrections. In this section we extend Appendix B of~\cite{Balasubramanian:2005qu} to general $J_{L,R}\gtrsim O(\sqrt{N})$. Before proceeding to our extension we warm up with a review of their original
calculation of $d_{J_L,J_R}$, valid in the regime where $J_L+J_R\sim
O(N)$ and $J_L-J_R \sim 0$. Our presentation is largely self-contained but
the reader may also find it useful to consult their more detailed
exposition.\footnote{We use a slightly different notation than
  \cite{Balasubramanian:2005qu}, who study the linear the combinations $(-J_+,J_-) \equiv  (J,\tJ)$. These linear combinations appear naturally in the gravitational description of the D1-D5 microstates as their angular momenta in the transverse $(x_1,x_2)$ and $(x_3,x_4)$ planes~\cite{Lunin:2002iz}.}

Appendix
B of~\cite{Balasubramanian:2005qu} computes the density of states for
$J_+\sim O(N)$ by starting with the partition function of the D1-D5 CFT at the orbifold
point with a chemical potential $\mu_+$,
\begin{equation}
\label{eq:Balasubramanian:2005quz}
Z(\beta,\mu_+) = \tr \, e^{-\beta(N-\mu_+ J_+)}\,.
\end{equation}
When $N$ is large this can be evaluated via the saddle point approximation, leading to an expression for the entropy at
fixed $J_+$. Note that in this expression, unlike in~\eqref{eq:rhoBH},
$\beta$
is not the inverse of the physical black hole temperature but rather
an auxiliary chemical potential conjugate to the left-moving
excitation number $N\equiv N_L$ (i.e. it is the inverse ``effective
temperature'' $\beta_L$ of the left-movers); in particular, the physical 
Hawking temperature $T_H=\left(T_L^{-1}+T_R^{-1}\right)^{-1}=0$.

The seed of the orbifold has $N_B$ left-moving
bosons and $N_F$ left-moving fermions. Out of these states, $n_B$ of
the bosons have $J_+ = 1$ and another $n_B$ have $J_+ =
-1$. Similarly, there are $n_F$ fermions each with $J_+ = \pm
1/2$. We will follow~\cite{Balasubramanian:2005qu} in leaving the numbers of species general to
faciliate comparison with their expressions.\footnote{When $M_4 = T^4$ the theory has $N_B = N_F = 8$, $n_B =1$ and $n_F =
4$. When $M_4 = K3$ the theory has $N_B = 24$, $n_B=1$, $N_F = n_F =
0$.} With these charge
assigments, the partition function~\eqref{eq:Balasubramanian:2005quz} is~\cite{Balasubramanian:2005qu} 
\begin{align}
Z(\beta,\mu_+) &= \tr\ e^{-\beta(N-\mu_+ J_+)}\nonumber\\
&= \prod_{n=1}^{\infty} \frac{\left[(1+z^{1/2}q^n)(1+z^{-1/2}q^{n})\right]^{n_F}(1+q^n)^{N_F-2n_F}}{\left[(1-zq^n)(1-z^{-1}q^n)\right]^{n_B}(1-q^n)^{N_B-2n_B}}\nonumber\\
&\equiv Z_B Z_F
\end{align}
where $q=e^{2\pi i \tau}=e^{-\beta}$ and $z=e^{2\pi i \nu} = e^{\beta\mu_+}$. 
This can be rewritten in terms of special functions as
\begin{equation}
Z_B=2^{n_B} q^{N_B/24} \eta(\tau)^{-N_B+3n_B}  \left[\frac{\sin(\pi\nu)}{\theta_1(\pi\nu|\tau)}\right]^{n_B}
\end{equation}
and
\begin{equation}
Z_F=2^{-N_F/2} q^{-N_F/24} \eta(\tau)^{-N_F/2}  \left[\frac{\theta_2(\pi\nu/2|\tau)}{\cos(\pi\nu/2)}\right]^{n_F} \theta_2(0|\tau)^{N_F/2-n_F}\,.
\end{equation}
In the thermodynamic limit $\beta\rightarrow 0$, corresponding to $N\gg
1$, the partition function simplifies:
\begin{equation}
Z(\beta,\mu_+) \approx \beta^{N_B/2} \left(\frac{\mu_+}{\sin{\pi\mu_+}}\right)^{n_B}  e^{\pi^2 c_s/6\beta}\,.
\end{equation}
Here $c_s$ is the central charge of the seed of the orbifold,
$c_s = N_B+ N_F/2$, which is $O(1)$. The first factor is the least
singular as $\beta\rightarrow 0$ and can be neglected.

The saddle
point approximation leads to the thermodynamic relations
\begin{equation}
\label{thermos}
N = -\left(\frac{\partial \log Z}{\partial \beta}\right)_{\beta\mu_+} = \frac{c_s\pi^2}{6\beta^2}+\frac{n_B \mu_+}{\beta} g(\mu_+),\quad\quad J_+= \left(\frac{\partial \log Z}{\partial (\beta\mu_+)}\right)_{\beta} =\frac{n_B}{\beta}g(\mu_+)\,,
\end{equation}
where
\begin{equation}
g(x) = \frac{1}{x} - \frac{\pi}{\tan\pi x}\,.
\end{equation}
From these expressions one obtains the entropy
\begin{eqnarray}
\label{Balasubramanian:2005quent}
S &=& \beta(N-\mu_+ J_+) + \log Z = \frac{c_s\pi^2}{3\beta}+n_B\log\left(\frac{\mu_+}{\sin{\pi\mu_+}}\right)\nn\\
&=& 2\pi\sqrt{\frac{c_s}{6}(N-\mu_+ J_+)}+\log\left(\frac{\mu_+}{\sin{\pi\mu_+}}\right)
\end{eqnarray}
and the expectation value of $J_+$,
\begin{equation}
\label{J}
J_+ = \frac{3 \mu_+ g(\mu_+)^2}{c_s\pi^2}\left[\sqrt{1+\frac{2c_s N\pi^2}{3\mu_+^2g(\mu_+)^2}}-1\right].
\end{equation}
In the last two expressions we used $n_B=1$. The magnitude of $J_+$ is controlled by the behavior of
$g(\mu_+)$. In~\cite{Balasubramanian:2005qu} $\mu_+$ is determined as follows: first, since
$N-\mu_+ J_+$ is not a positive operator when $|\mu_+| > 1$, we must
restrict to $|\mu_+|\leq 1$ in order to obtain a well-defined partition function.\footnote{This bound can also be obtained
  by requiring that the density of states is real.} Next, one demands
that $J_+ \sim O(N)$. As $\mu_+\rightarrow 0$, $g$ vanishes linearly,
\begin{equation}
g(\mu_+) = \frac{\pi^2}{3}\mu_+ + O(\mu_+^3),
\end{equation}
but it diverges as $\mu_+\rightarrow \pm 1$:

\begin{equation}
g(\mu_+) = \frac{-1}{\mu_+ \mp1}\pm 1 + O(\mu_+\mp 1).
\end{equation}
Thus if we take $|\mu_+ - \text{sign}(J_+)|\sim N^{-1/2}$, \eqref{J}
implies $J_+\sim O(N)$ and we have the desired
enhancement. Making this choice in~\eqref{Balasubramanian:2005quent}, the density of states in this regime is
\begin{equation}
\label{dos1j}
d_{J_+,J_-}= e^{ 2\pi\sqrt{\frac{c_s}{6}(N-|J_+|)}} \quad\quad \text{when } \quad\quad J_+\sim O(N), \ J_- \sim 0
\end{equation}
up to subleading corrections in $N$. This completes our review of~\cite{Balasubramanian:2005qu}.

To study the density of states at general $J_{L/R}$ we must also
turn on a chemical potential for~$J_-$:

\begin{align}
\label{Balasubramanian:2005qutwojs}
Z(\beta,\mu_+,\mu_-) &= \tr\ e^{-\beta(N-\mu_+ J_+ -\mu_- J_-)}\nonumber\\
&\hspace{-1.8cm}= \prod_{n=1}^{\infty} \frac{\left[(1+z^{1/2}q^n) (1+z^{-1/2}q^n) (1+\tilde{z}^{1/2}q^{n}) (1+\tilde{z}^{-1/2}q^{n})\right]^{n_F}(1+q^n)^{N_F-4n_F}}{\left[(1-z q^n) (1-z^{-1} q^n) (1-{\tilde z} q^n) (1-{\tilde z}^{-1}q^n)\right]^{n_B}(1-q^n)^{N_B-4n_B}}\nonumber\\
&\hspace{-1.8cm}\equiv Z_B Z_F,
\end{align}
where $q=e^{2\pi i \tau}=e^{-\beta}$, $z=e^{2\pi i \nu} =
e^{\beta\mu_+}$ and $\tilde z=e^{2\pi i \tilde \nu} =
e^{\beta\mu_-}$. The second line follows from the fact that the theory
has $n_B$ states with charges $(J_+,J_-) = (\pm 1,0), (0,\pm 1)$, and similarly for the fermions. We find

\begin{align}
Z_B &= 2^{2n_B} q^{N_B/24} \eta(\tau)^{-N_B+6n_B} \left[\frac{\sin(\pi\nu_+)}{\theta_1(\pi\nu_+|\tau)}\right]^{n_B}\left[\frac{\sin(\pi\nu_-)}{\theta_1(\pi\nu_-|\tau)}\right]^{n_B}
\end{align}
and
\begin{align}
Z_F &= 2^{-N_F/2} q^{-N_F/24} \eta(\tau)^{-N_F/2} \left[\frac{\theta_2(\pi\nu_+/2|\tau)}{\cos(\pi\nu_+/2)}\right]^{n_F} \left[\frac{\theta_2(\pi\nu_-/2|\tau)}{\cos(\pi\nu_-/2)}\right]^{n_F} \theta_2(0|\tau)^{N_F/2-2n_F}.
\end{align}

In the thermodynamic limit $\beta\rightarrow 0$ corresponding to large
$N$, the partition function once again simplifies:
\begin{equation}
Z(\beta,\mu_+,\mu_-) \approx \beta^{N_B/2} \left(\frac{\mu_+}{\sin{\pi\mu_-}}\right)^{n_B} \left(\frac{\mu_+}{\sin{\pi\mu_-}}\right)^{n_B} e^{\pi^2 c_s/6\beta}.
\end{equation}
The saddle point analysis leads to
\begin{align}
\label{thermos2}
N &= -\left(\frac{\partial \log Z}{\partial \beta}\right)_{\beta\mu_+,\beta\mu_-} = \frac{c_s\pi^2}{6\beta^2}+\frac{n_B \mu_+}{\beta} g(\mu_+) + \frac{n_B \mu_-}{\beta} g(\mu_-),\nonumber\\
J_+&= \left(\frac{\partial \log Z}{\partial (\beta\mu_+)}\right)_{\beta,\beta\mu_-} =\frac{n_B}{\beta}g(\mu_+),\quad\quad J_-= \left(\frac{\partial \log Z}{\partial (\beta\mu_-)}\right)_{\beta,\beta\mu_+} =\frac{n_B}{\beta}g(\mu_-),
\end{align}
and the entropy
\begin{align}
S &= \beta(N-\mu_+ J_+ - \mu_- J_-) + \log Z = \frac{c_s\pi^2}{3\beta}+n_B\log\left(\frac{\mu_+}{\sin{\pi\mu_+}}\right) +n_B\log\left(\frac{\mu_-}{\sin{\pi\mu_-}}\right)\nn\\
&= 2\pi\sqrt{\frac{c_s}{6}(N - \mu_+ J_+-\mu_- J_-)}+\log\left(\frac{\mu_+}{\sin{\pi\mu_+}}\right) +\log\left(\frac{\mu_-}{\sin{\pi\mu_-}}\right)
\end{align}
where we again used $n_B=1$. Similarly, one finds the R-charges

\begin{equation}
\label{jlr}
J_{\pm} = \frac{3f_{\pm}}{c_s\pi^2}\left[\sqrt{1+\frac{2c_sN\pi^2 g(\mu_{\pm})^2}{3f_{\pm}^2}}-1\right]
\end{equation}
where
\begin{equation}
f_{\pm}= g(\mu_{\pm})\left[\mu_{\pm} \ g(\mu_{\pm})+ \mu_{\mp} \ g(\mu_{\mp})\right].
\end{equation}
As above we determine $\mu_{\pm}$ in the density of states at
$J_{\pm}\sim O(N)$ by requiring that~\eqref{jlr} implies $J_{\pm}\sim O(N)$. Once again this implies $|\mu_{\pm} - {\rm sign} (J_{\pm})|\sim N^{-1/2}$
and so
\begin{equation}
\label{dos2js}
d_{J_L,J_R} = e^{ 2\pi\sqrt{\frac{c_s}{6}(N-|J_+|-|J_-|)}} \quad\quad \text{when } \quad\quad (J_+,J_-)\sim O({N}).
\end{equation}
This is the expression we used to evaluate $\chi$ in~\eqref{eq:chijtj}.

So far we have computed the density of states when $J_{L/R}\sim O(N)$. However, typical microstates of the $M=0$ BTZ black hole have $J_{L/R} \sim O(\sqrt{N})$. When $J_{L/R}\sim O(\sqrt{N})$, we find that the density of states is

\begin{equation}
\label{dos2jstyp}
d_{J_L,J_R} = e^{ 2\pi\sqrt{\frac{c_s}{6}\left[N-\gamma_+(J_+)|J_+|-\gamma_-(J_-)|J_-|\right]}}   \quad\quad \text{when } \quad\quad      (J_+,J_-)\sim O(\sqrt{N}).
\end{equation}
We have used time-reversal invariance to fix $\mu_\pm =\gamma_\pm\   \text{sign}(J_\pm)$ for some $0<\gamma_\pm<1$, but the $\gamma_\pm(J_\pm)$ must be determined numerically by inverting the transcendental relation~\eqref{jlr}. We will now quantify the error in our computation of $\chi$ in~\eqref{eq:chijtj} caused by our use of~\eqref{dos2js} over the whole range of $J_{L/R}$.\footnote{One might have worried about logarithmic corrections to the density of states, but they are subleading compared to modifications of the form \eqref{dos2jstyp}.}

We should have used~\eqref{dos2jstyp} instead of~\eqref{dos2js} in the appropriate region of the integral in~\eqref{eq:chijtj} when we computed $\chi$. Consider the fractional error we made in computing $\chi$,

\be
\label{fracdiff}
\frac{\Delta \chi}{\chi}  = \frac{ \int_{-N}^N dJ_+\ dJ_-\ \left[
e^{ 2\pi\sqrt{\frac{c_s}{6}\left[N-\gamma_+(J_+)|J_+|-\gamma_-(J_-)|J_-|\right]}} - e^{ 2\pi\sqrt{\frac{c_s}{6}\left(N-|J_+|-|J_-|\right)}}
\right]   \left(J_L^2+J_R^2\right)   
} {
\int_{-N}^N dJ_+\ dJ_-\ e^{ 2\pi\sqrt{\frac{c_s}{6}\left(N-|J_+|-|J_-|\right)}}\left(J_L^2+J_R^2\right)   }.
\ee
To be complete we have integrated over the entire range of $J_\pm$, though the contribution from $J_\pm \sim O(N)$ vanishes since~\eqref{dos2js} approaches~\eqref{dos2jstyp} in that regime. Our conclusion in the main text that $\alpha\sim O(c^0)$ will hold true so long as the fractional error does not scale with $N$.

\begin{figure}[ht!]
 \begin{center}
  \includegraphics[width=0.45\textwidth]{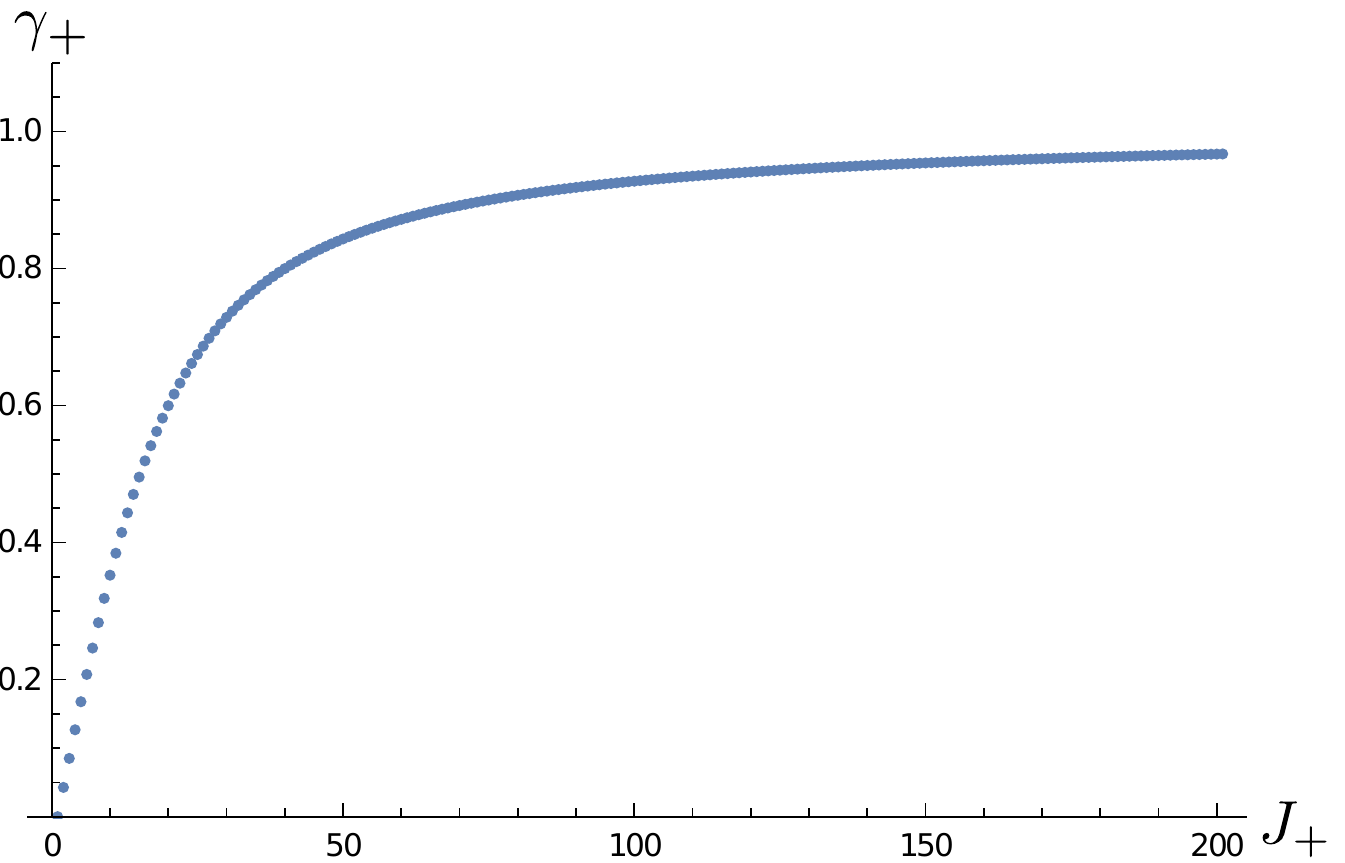} \hspace{0.5cm} \includegraphics[width=0.45\textwidth]{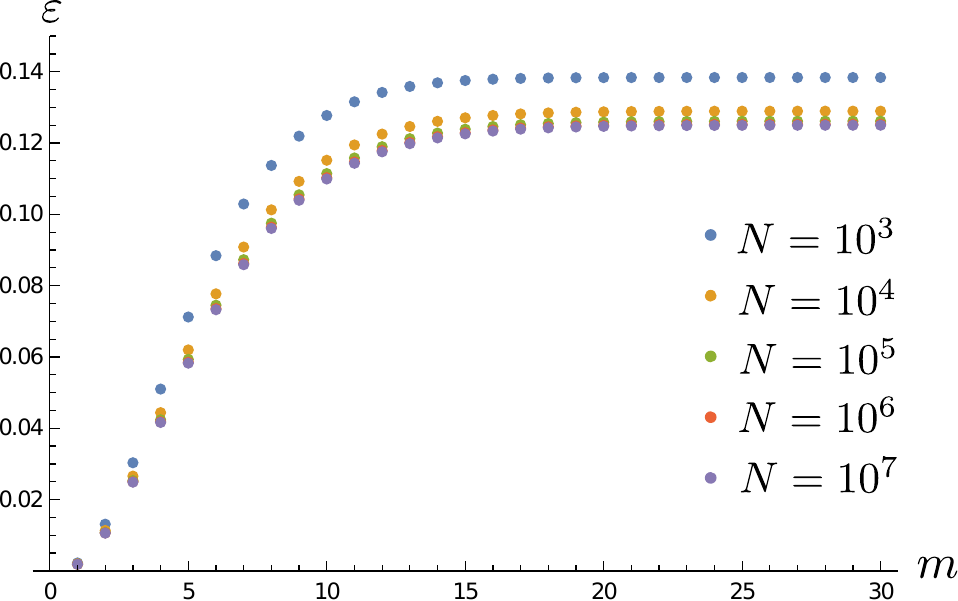}
  \end{center}
  \caption{Left: $\gamma_+(J_+)$ for $N=1000$.\hspace{1cm} Right: $\varepsilon(m)$ at large $N$.}
  \label{fig:vareps}
 \label{fig:gammaofJ}
 \end{figure}

We will analyze the simpler quantity

\be
\varepsilon(m) \equiv \frac{\int_{0}^{m\sqrt{N}} dJ_+\ \left[e^{\sqrt{N-\gamma_+(J_+)|J_+|}} - e^{\sqrt{N-|J_+|}}\right] \ J_+^2} {\int_{0}^{N} dJ_+\ e^{\sqrt{N-|J_+|}} \ J_+^2}.
\ee
This quantity misses contributions from $|J_+|~>~m\sqrt{N}$ but such contributions will be negligible when $m$ is sufficiently large. Here $\gamma_+\equiv \mu_+\ \text{sign}(J_+)$, obtained by inverting~\eqref{J}, is plotted in the left panel of figure~\ref{fig:gammaofJ}. In the large-$N$ limit and for sufficiently large $m$, $\varepsilon$ is equal to the fractional error~\eqref{fracdiff} up to an overall $O(1)$ multiplicative factor (see the remarks in footnote~\ref{jfootnote}).

In the right panel of figure~\ref{fig:vareps} we plot $\varepsilon(m)$ numerically for several different values of $N$. There are two salient features. The first is the rapid convergence in $m$ to an asymptotic value $\varepsilon_\star$: we do not need to take $m$ very large in order to obtain a good approximation of $\Delta \chi/\chi$, once the $O(1)$ multiplicative factor is restored. This follows from the fact that $\gamma_+$ is monotonic and approaches 1 quite rapidly above $J_+\sim O(\sqrt{N})$. The second feature is the crucial one: convergence $\varepsilon_\star\rightarrow 0.1249$ at large $N$. This leads to our conclusion that $\alpha\sim O(c^0)$.

\bibliographystyle{toine}
\bibliography{references}

\end{document}